# Microelectromagnets for the manipulation of biological systems

Hakho Lee*, Alfreda M. Purdon*,[#] and Robert M. Westervelt*,[§]

* Department of Physics, Harvard University, Cambridge, Massachusetts 02138.

[#] Department of Chemistry, Harvard University, Cambridge, Massachusetts 02138.

[§] Division of Engineering and Applied Sciences, Harvard University, Cambridge, Massachusetts 02138.

## Abstract

Microelectromagnet devices, a ring trap and a matrix, were developed for the microscopic control of biological systems. The ring trap is a circular Au wire with an insulator on top. The matrix has two arrays of straight Au wires, one array perpendicular to the other, that are separated and topped by insulating layers. Microelectromagnets can produce strong magnetic fields to stably manipulate magnetically tagged biological systems in a fluid. Moreover, by controlling the currents flowing through the wires, a microelectromagnet matrix can move a peak in the magnetic field magnitude continuously over the surface of the device, generate multiple peaks simultaneously and control them independently. These capabilities of a matrix can be used to trap, continuously transport, assemble, separate and sort biological samples on micrometer length scales. Combining microelectromagnets with microfluidic systems, chip-based experimental systems can be realized for novel applications in biological and biomedical studies.

**INTRODUCTION**

The development of technology for manipulating biological cells, proteins, and molecules has been important in cell and molecular biology, and in medicine, advancing scientific research and clinical applications (Mehta et al., 1999; Bustamante et al., 2000). The controlled synthesis of uniform, micrometer-size polymer beads that can be attached to biological systems and the manipulation of these bead-bound objects have introduced new diagnostic and therapeutic procedures both *in vivo* and *in vitro*: fluorescent microspheres have been used to label cells and facilitate cell separation (Rembaum and Dreyer, 1980), polymer beads have been manipulated with optical tweezers (Ashkin, 1997) to characterize the mechanical properties of biological systems such as DNA (Perkins et al., 1994), flagella motor of *E-coli* (Ryu et al., 2000), and red blood cells (Svoboda et al., 1992). Magnetic beads have also played an important role in biological and biomedical studies (Häfeli, 1997). By functionalizing its surface with antibodies, peptides or lectins, a magnetic bead can form a highly specific link to a target system, permitting fast and easy manipulation of the bead-bound object with external magnetic fields. The elastic properties of DNA (Bustamante et al., 1994) and the rotation of DNA during transcription by RNA polymerase (Harada et al., 2001) have been studied using magnetically tagged DNA molecules. Magnetic separation and assay of cells and proteins have become well-established methods in biomedical research (Radbruch et al., 1994).

Using macroscopic external magnets, either permanent magnets or electromagnets, as the sources of the magnetic field, various magnetic tweezers have been implemented to manipulate magnetic beads in a fluid (Ziemann et al., 1994; Haber and Wirtz, 2000;

Gosse and Croquette, 2002). However, due to their mechanical complexity and macroscopic size, these devices cannot be easily miniaturized and integrated with microfluidic systems, which limits their manipulation capabilities. To overcome the drawback, chip-based magnetic tweezers and separators, which use lithographically patterned planar coils or wires to generate magnetic fields, were developed and controlled magnetic beads in a small volume of sample solution (Choi et al., 2001; Deng et al., 2001). Still, the manipulation capability of these devices is limited; the magnetic field intensity and direction can be changed but the spatial pattern is fixed, making it difficult to perform a simultaneous and independent manipulation of single or multiple biological samples. To provide sophisticated manipulation functions required in biological or biomedical experiments, new magnetic manipulation systems that can control single or multiple magnetic beads on microscopic length scales have to be developed.

In this paper we report the development and the novel operation of microelectromagnets as a new versatile manipulation tool for biological experiments. Microelectromagnets, consisting of multiple layers of lithographically defined conducting wires, can generate strong magnetic field patterns over microscopic length scales. Previously, microelectromagnets were used to control the motion of cold atoms in a vacuum (Johnson et al., 1998; Dekker et al., 2000) and to manipulate magnetic particles in a fluid (Lee et al., 2001). Here, we describe the noninvasive and microscopic manipulation of single or multiple biological cells in a fluid using two types of microelectromagnets, a ring trap and a matrix. A ring trap is a circular wire topped with an insulating layer; a matrix is two layers of straight wires aligned perpendicular to each other and covered with insulating layers. Yeast (*Saccharnomyces cerevisiae*) was chosen as a target sample to demonstrate a general approach to the micromanipulation of non-magnetic biological systems with microelectromagnets. A protocol was developed to attach magnetic beads to yeast cells, which can also be applied to other biological systems. In the experiments shown here, a single yeast cell was trapped at a given location with a ring trap, and the matrix was used to individually manipulate a number of yeast cells at different locations.

The microelectromagnet matrix has a noticeable advantage over other magnetic manipulation tools, making it possible to conduct biological experiments on micrometer length scales. By controlling a current flow in each wire, a matrix can produce desired magnetic field patterns without any modification in the structure of the device. For example, a matrix can create and move a single peak in the magnetic field magnitude, precisely positioning a single, magnetically tagged biological system with micrometer scale resolution. Using the same device, multiple magnetic field peaks can be created and controlled to manipulate multiple biological samples simultaneously and independently. Two different cells, for example, can be brought together to force interactions between them, or they can be sorted to desired locations. Because the matrix does not require external magnets to operate, it reduces the complexity and the size of experimental setup. Furthermore, microelectromagnets can be easily mass-produced using lithographic techniques and integrated into microfluidic systems, realizing a lab-on-a-chip system for microanalysis (Harrison et al., 1993; Fu et al., 1999).

## MATERIALS AND METHODS

### Microelectromagnet fabrication

The microelectromagnets were fabricated using standard optical lithography and metal lift-off technique on silicon/silicon oxide substrates. For the conducting wires, Cr (thickness 10 nm) was deposited, followed by Au deposition (thickness 400 nm) by thermal evaporation. Polyimide (PI 2556, HD Microsystems, Parlin, NJ) was spin-coated on top of the wires and cured to form insulating layers. Benzocyclobutene (Cyclotene 3022-35, Dow Chemical, MI) that has a good planarization property, was spin-coated on top of each Polyimide layer and cured to make the surface of the device flat.

### Fluidic system

A fluidic chamber was separately fabricated with polydimethylsiloxanes (Sylgard 184, Dow Corning, MI) using soft lithography technique (Whitesides et al., 2001). Microelectromagnets and the fluidic chamber were treated with $O_2$ plasma to render their surfaces hydrophilic, followed by the sealing of the fluidic chamber on top of the microelectromagnets. Fluidic connection was made by silicone tubing. Sample solution containing target biological systems was introduced into the fluidic chamber over the microelectromagnets by a low flow rate (5 μl/min) peristaltic pump (VWR, Bristol, CT).

**System setup**

An upright microscope was built using Olympus BHM-2 microscope parts (Olympus America Inc., Melville, NY). Epifluorescent excitation was provided by an Ar ion laser (177-G02, Spectra Physics, Mountain View, CA) through a 515NB3 filter (Omega Optical Inc., Brattleboro, VT). The emitted light from the sample was separated from the background with a 650CFLP long pass filter (Omega Optical Inc.) and captured by a CCD camera (CoolSNAP FX, Roper Scientific, Tucson, AZ). The microelectromagnet device was mounted on a copper stage with a thermoelectric cooler for temperature control. By keeping the temperature of the stage at 4°C, the temperature of the microelectromagnets could be maintained around 25°C, which was measured indirectly by monitoring the change of electrical resistance of microelectromagnets. Currents to the microelectromagnet were provided by a custom-made, computer-controlled current supply with 20 output channels.

**Magnetic bead preparation**

Magnetic beads, diameter 2.8 μm and coated with a polyurenthane layer (Dynabeads M-280, Dynal Biotech, Lake Success, NY), were triple washed with phosphate buffered saline (PBS) solution and re-suspended in PBS solution ($2.0 \times 10^9$ bead/ml). Concanavalin–A (Con–A) solution was prepared by suspending the powder form of the protein (Sigma–Aldrich, St. Louis, MO) in PBS solution (5 mg/ml). Magnetic bead solution (0.5 ml) and Con–A solution (1 ml) were mixed and incubated for 24 hours at 20 °C to coat Con–A on magnetic beads.

**Magnetic bead binding to yeast**

Dried baker's yeast (*Saccharnomyces cerevisiae*) was suspended in Tris(hydroxymethyl)aminomethane (Tris) solution buffered at pH = 7.3 and incubated for 1 hour at 30 °C. Con–A coated bead solution (5 μl) and yeast suspension (30 μl) were then mixed with binding PBS buffer (40 μl) which contains 5 mM $Ca^{2+}$, 5 mM $Mn^{2+}$, 0.1 M NaCl and 5 mM EDTA (Sharon and Lis, 1972). The mixture was incubated in rotator for 24 hours at 20 °C, followed by triple washing with distilled water. The final sample was suspended in PBS buffer (10 ml). Two dyes, FUN 1 and Calcofluor (Molecular Probes Inc., Eugene, OR) were added to the final solution to stain yeast cells.

## RESULTS AND DISCUSSION

### Device structure

Fig. 1 shows schematic diagrams and micrographs of two specific types of microelectromagnet that were fabricated and tested: a ring trap (Fig. 1, *A–C*) and a matrix (Fig. 1, *D–F*). The ring trap is a current-carrying circular wire covered with an insulating layer. The matrix consists of two sets of straight conducting wires aligned perpendicular to each other, separated by insulating layers, with additional insulating layers fabricated on top. The conducting wires can be patterned either by optical lithography or by electron-beam lithography, depending on the diameter *a* (ring trap) and the wire pitch *p* (matrix) required for experiments. For a single cell trapping, the diameter and the wire pitch should be comparable to the size of the target cell. The insulating layers are necessary to prevent electrical shorting between wires and samples suspended in a fluid and to provide flat surface for sample manipulations. Furthermore, as illustrated in Fig. 1 *B*, the thickness of insulting layers (*d*) has to be controlled to avoid the creation of the magnetic field maximum right on top of wires. From numerical calculation, the criteria of insulator thickness, $d \geq 0.32a$ and $d \geq 3p/(2^{3/2}\pi)$, were obtained for a ring trap and a matrix, respectively.

### Principle of magnetic trapping

The operating principle of magnetic manipulation with a microelectromagnet is illustrated in Fig. 1 *A*. The magnetic field generated by a current-carrying wire in a microelectromagnet pulls magnetic objects suspended in a fluid toward maxima in the field magnitude, trapping them on the surface of an insulating layer. By moving the position of the magnetic field maximum, trapped objects can also be moved over the surface of the device. With the insulator thickness $d$, the trapping magnetic field magnitude on the surface of the device is $B \propto JA/d$, where $J$ is the current density in the wire and $A$ is the cross sectional area of the wire. For a given device, the current density in the wire can be increased by cooling down the wire. For microelectromagnets reported here, which have $A \sim 1.6$ μm$^2$ and $d \sim 2.0$ μm, current densities as high as $J \sim 0.5$ A/μm$^2$ were achieved at room temperature, producing magnetic field magnitudes up to $B \sim 0.1$ T. To prevent thermal breakdown from Joule heating, and electromigration, the device was placed on a copper stage cooled by a thermoelectric cooler.

Because magnetic objects are suspended in a fluid, they undergo thermal motions as well as being pulled by a trap. The potential energy of a trapped object is $U = -mB$, where $m$ is the object's magnetic moment. In thermal equilibrium, the chemical potentials of magnetic objects inside and outside the trap are equal to each other (Kittel and Kroemer, 1980), which determines the average number of objects per unit volume inside the trap $n_T = n_O \exp(|U|/k_B T)$, where $n_O$ is the number of objects per unit volume outside the trap, $k_B$ is Boltzmann's constant and $T$ is the temperature. In a single trap with trapping volume $V_T$, magnetic objects will be trapped provided $n_O V_T \geq \exp(-|U|/k_B T)$. This gives the criterion for the magnetic moment of an object $m \geq -(k_B T/B)\log n_O V_T$, which is required for stable trapping at a given temperature. Another important factor for stable manipulation is the residence time of a trapped object inside a trap. The capture rate of a magnetic object into a trap with trapping volume $V_T$ is $k_C \sim n_O v V_T^{2/3}$, where $v$ is the mean speed of the object in thermal equilibrium. From the principle of detailed balance, the escape rate of a trapped object is then $k_E = k_C n_O/n_T = k_C \exp(-|U|/k_B T)$. The mean residence time of a trapped object inside a trap is then $\tau = k_E^{-1} = k_C^{-1} \exp(|U|/k_B T)$.

Using $n_O = 10^{-7}/\mu m^3$, $V_T = 25$ $\mu m^3$ and $B = 0.1$ T for the microelectromagnets reported here, the minimum magnetic moment for stable trapping at $T = 25°C$ is $m = 5\times10^{-19}$ Am$^2$. In an external magnetic field, the saturation magnetic moments of magnetic beads (for example, Dynabeads M-280, Dynal Biotech) are $\sim 10^{-14}$ Am$^2$, far exceeding the required minimum magnetic moment. Once trapped, the residence time of the beads inside a trap is $\gg 10^5$ sec with $B = 0.1$ T at $T = 25°C$. This allows stable trapping and manipulation of the magnetic beads as well as magnetic bead bound biological systems at room temperature with microelectromagnets.

**Magnetic fields generation with a microelectromagnet matrix**

While the ring trap can produce a fixed magnetic field pattern, the microelectromagnet matrix, consisting of two perpendicular arrays of wires, can create complex magnetic field patterns by controlling the individual currents in each wire. Fig. 2 shows the examples of the magnetic field profiles calculated for a microelectromagnet matrix with 16 wires in each conducting layer (a 16×16 wire matrix). The wire pitch and the total thickness of insulator were 8 μm and 4 μm, respectively and the magnetic field magnitude was computed on the surface of the device.

Fig. 2 *A* shows a single peak in the magnetic field magnitude with the current distribution in wires. The current in each wire was optimized by the least square

algorithm to produce a Gaussian shape peak ($B = 10$ mT) in the field magnitude. A set of current distributions was then calculated to move the single peak across the matrix in steps less than the wire pitch (Fig. 2 *B*). This profile can be used to trap and precisely position the target sample at desired locations. The same optimization process was applied to determine the current distributions for the simultaneous and independent control of multiple peaks. In Fig. 2 *C*, two separate peaks were created with one peak fixed and the other moving independently. Using this profile, two different biological targets can be separated for sorting or they can be brought together to force interactions. Fig. 2 *D* shows how four peaks can trap multiple objects and bring them simultaneously to one position. By reversing the sequence, a single group of objects can be split into multiple groups that can be controlled independently. Fig. 2 *E* shows two peaks, one fixed at the center and the other circling around, that can be used to rotate or twist target samples. These examples demonstrate the advantage of a matrix as a versatile manipulation tool: by adjusting the currents in wires, the matrix can create desired magnetic field patterns to meet specific experimental purposes without changing the structure of the device.

**Cell trapping with a ring trap**

Biological system manipulation with microelectromagnets was performed using baker's yeast (*Saccharnomyces cerevisiae*) as a target sample. Fig. 3 *A* shows a scanning electron microscope image of a yeast cell attached to a magnetic bead of diameter 2.8 µm (M–280, Dynal Biotech). The surface of the magnetic bead was functionalized with Concanavalin–A, a lectin (Sharon and Lis, 1972) that recognizes sugar molecules ($\alpha$-D-mannose) on the wall of the yeast cell and makes specific binding to them with the binding force ~ 100 pN (Neumann et al., 2002). The saturation magnetic moment of the bead is ~ $3\times10^{-14}$ $Am^2$, which is sufficient for stable micromanipulation of the yeast cell in a fluid with microelectromagnets.

A two-color fluorescent dye (FUN 1; Molecular Probes), which utilizes the metabolic differences between viable and non-viable cells for different color labeling, was used to determine the viability of the yeast cells in the final sample solution. Additional dye (Calcofluor; Molecular Probes) was added to stain cell walls for better imaging. Fig. 3 *B* shows a fluorescent microscope image of stained yeast samples with magnetic beads. Viable yeast cells were observed red with vivid cylindrical inner structures, while non-viable cells were diffusively green fluorescent (Fig. 3 *B* inset).

The sequence of images in Fig. 3 *C* shows the operation of a ring trap ($a$ = 5 µm) to trap a single, viable yeast cell. Solution containing the stained yeast samples was introduced into a fluidic chamber placed over the ring trap with the magnetic field off. With the current $I$ = 50 mA, a magnetic field peak ($B$ = 6 mT) was produced at the center of the ring above the insulating layer. The temperature of the device, indirectly measured by monitoring the change of electrical resistance of the trap, was 25°C with the active cooling by a thermoelectric cooler attached on the back of the device. Several seconds after the magnetic field was on, a yeast cell bound to a magnetic bead was trapped. Because the size of the trap was comparable to that of yeast, only a single cell was trapped. With the magnetic field on, the trapped yeast cell remained stationary due to the large magnetic moment of the bead. Furthermore, the cell remained viable throughout the experiment, demonstrating the non-invasive, biocompatible manipulation capability of the microelectromagnets.

**Cell manipulation with a matrix**

More dynamic and versatile micromanipulation of biological systems was performed with a microelectromagnet matrix. Fig. 4 shows the control of yeast cells by a 10×10 wire matrix with the wire pitch 8 μm and the insulator thickness 4 μm. Currents in all wires were adjusted to create desired magnetic field patterns with the maximum $B = 40$ mT. Currents were supplied by 20 current sources, one for each wire, that were individually controlled by a computer. A thermoelectric cooler on the back of the device kept the temperature of the device around 25°C.

Fig. 4 *A* shows the operation of the matrix to trap and continuously move a single yeast cell bound to a magnetic bead on the surface of the device. The trapping magnetic peak was generated and moved by increments smaller than the wire pitch as shown in Fig. 2 *A*, resulting in the continuous transport of the trapped yeast. Multiple samples were also manipulated as shown in Fig. 4 *B*. Two groups of samples, a single cell with three magnetic particles and two cells with a magnetic particle, were trapped at two different locations. Subsequently, while the single cell was held still, the group of two yeast cells was moved using the magnetic field patterns shown in Fig. 2 *B*. These operations show how the matrix can be a sophisticated manipulation tool for biological or biomedical studies. Optimized magnetic field patterns can be created to meet specific experimental needs. For example, a matrix can trap, move and position biological systems at desired locations where further characterizations can be performed; multiple samples can be controlled simultaneously and independently, allowing sorting and separation of specific samples. Furthermore, the matrix could be used to assemble various biological systems into predetermined patterns.

As an example for biological applications, Fig. 5 shows a cell-sorting operation on micrometer length scales with a matrix. A group of yeast cells, two nonviable cells and one viable cell, was initially trapped by creating a single magnetic field peak (Fig. 5 *A*). Subsequently, the single peak was split into two smaller peaks which were controlled independently: while one of the peaks was holding the non-viable cells, the other peak was moved away to separate the viable cell from the rest of the group (Fig. 5). Reversing the sequences, two different cells can be converged into one position to assay their interactions. By simply adjusting currents in wires, more peaks can be added and controlled simultaneously depending on the number of samples to be manipulated.

**Cell rotation with a time-varying magnetic field**

Besides creating static magnetic fields, a microelectromagnet matrix can generate dynamic field patterns with time-varying currents. This capability, together with the magnetic polarity of the magnetic beads, can be used to rotate specific target samples on micrometer length scales. As an example, Fig. 6 shows the rotation of yeast cells with the 10×10 wire matrix. Two sinusoidal currents at the same frequency *f* but with phase difference 90° were applied to two wires crossing each other (Fig. 6 *A*). The trapping magnetic field was formed at the crossing point; the direction of magnetic field on the surface is rotating at the same frequency. Fig. 6 *B* shows the initial trapping of the yeast sample at the crossing point. The sequence of images in Fig. 6 *C* shows one complete revolution of the sample with the maximum rotational frequency $f_M = 2$ Hz. For a spherical particle, $f_M$ can be estimated using Stokes formula (Reif, 1965). When a particle of diameter $a$ is moving at speed $u$ in a fluid of viscosity $\eta$, the viscous drag by the fluid on the particle is $3\pi\eta a u$. For a particle rotating at frequency $f$ with the distance $r$ away from the center of rotation, $u = 2\pi f r$ and the retarding torque by the fluid is $6\pi^2 \eta a r^2 f$. At steady state, the external torque by magnetic field is balanced with the retarding torque by the fluid. Because the maximum torque on a magnetic bead by a matrix is $mB$, $f_M = mB/6\pi^2 \eta a r^2$. Approximating the yeast cell as a sphere with $a = 5$ μm and using $\eta = 1.5 \times 10^{-3}$ kg/m/s for water, $m = 3 \times 10^{-14}$ Am$^{-2}$ for the magnetic particles and $B = 40$ mT by the matrix, the expected $f_M = 2.8$ Hz for the yeast sample shown here. This value is in good agreement with the observed result $f_M = 2$ Hz.

With time varying magnetic fields, a matrix can also twist magnetic beads attached to biological objects. By monitoring the response of the magnetic beads, the mechanical properties of target samples can be measured, enabling the magnetic twisting cytometry (Fabry et al., 2001) on micrometer length scales. In addition, due to their small size and geometry, a matrix can generate localized ac electromagnetic fields at radio and microwave frequencies along with static magnetic fields. With the static fields trapping objects, the ac fields can perturb and sense the responses of the objects for their characterization. These capabilities are well suited for possible applications in magnetic resonance imaging (MRI) in small length scales.

**CONCLUSIONS**

Various methods for microscopic control and manipulation of biological systems, including optical tweezers, patch clamps, dielectrophoretic traps and magnetic tweezers, have been developed and successfully used in many biological experiments (Neher and Sakmann, 1976; Ashkin, 1997; Morgan et al., 1999; Gosse and Croquette, 2002). In this paper, we have reported the development and the operation of microelectromagnets, especially the matrix, as a new tool for the micromanipulation of biological systems. We have demonstrated two-dimensional micromanipulations of yeast cells that are attached to magnetic beads: a single cell was trapped and transported continuously, multiple groups of cells were moved simultaneously or even rotated, and cell sorting operation was performed as an example of practical applications.

Several important aspects make a microelectromagnet matrix a useful manipulation tool for biological and biomedical applications. First, the matrix generates strong and localized magnetic fields on micrometer length scales, enabling stable manipulation of a single biological object in a fluid. The manipulation process is noninvasive, because of the biocompatibility of the magnetic field. Second, a target biological object can be trapped and moved continuously to specific locations, allowing precise positioning of the sample. Furthermore, by controlling currents, the same device can manipulate multiple biological samples independently. With this capability, the matrix can be used to perform various experimental and practical applications such as cell sorting, microassay, and cytometry on micrometer length scales. Third, the matrix is self-contained, operating without any external magnetic field sources or scanning instruments.

It can be readily integrated into a microscope stage, reducing the complexity and the size of experimental setup. Finally, because microelectromagnets are fabricated with existing microfabrication technology, they can be cost-effective and easily mass-produced. By incorporating microfluidic structures, a chip based analysis system can be realized (Quake and Scherer, 2000). Microelectromagnets, together with the well-established techniques for binding magnetic beads to biological systems, can be a powerful manipulation tool for a new class of biological and biomedical studies.

**ACKNOWLEDGEMENTS**

We thank X. Zhuang and M. Bawendi for their helpful comments. This work was supported by the Nanoscale Science and Engineering Center at Harvard under NSF grant PHY-0117795 and the ONR under grant N00014-99-1-0347.

Ashkin, A. 1997. Optical trapping and manipulation of neutral particles using lasers. *Proc. Natl. Acad. Sci. U.S.A.* 94:4853-4860.
Bustamante, C., J. C. Macosko, and G. J. Wuite. 2000. Grabbing the cat by the tail: manipulating molecules one by one. *Nat. Rev. Mol. Cell. Biol.* 1:130-136.
Bustamante, C., J. F. Marko, E. D. Siggia, and S. Smith. 1994. Entropic elasticity of lambda-phage DNA. *Science* 265:1599-1600.
Choi, J. W., T. M. Liakopoulos, and C. H. Ahn. 2001. An on-chip magnetic bead separator using spiral electromagnets with semi-encapsulated permalloy. *Biosens. Bioelectron.* 16:409-416.
Dekker, N. H., C. S. Lee, V. Lorent, J. H. Thywissen, S. P. Smith, M. Drndic, R. M. Westervelt, and M. Prentiss. 2000. Guiding neutral atoms on a chip. *Phys. Rev. Lett.* 84:1124-1127.
Deng, T., G. M. Whitesides, M. Radhakrishnan, G. Zabow, and M. Prentiss. 2001. Manipulation of magnetic microbeads in suspension using micromagnetic systems fabricated with soft lithography. *Appl. Phys. Lett.* 78:1775-1777.
Fabry, B., G. N. Maksym, J. P. Butler, M. Glogauer, D. Navajas, and J. J. Fredberg. 2001. Scaling the microrheology of living cells. *Phys. Rev. Lett.* 87:148102.
Fu, A. Y., C. Spence, A. Scherer, F. H. Arnold, and S. R. Quake. 1999. A microfabricated fluorescence-activated cell sorter. *Nat. Biotechnol.* 17:1109-1111.
Gosse, C., and V. Croquette. 2002. Magnetic tweezers: micromanipulation and force measurement at the molecular level. *Biophys. J.* 82:3314-3329.
Haber, C., and D. Wirtz. 2000. Magnetic tweezers for DNA micromanipulation. *Rev. Sci. Instrum.* 71:4561-4570.
Häfeli, U. 1997. *Scientific and clinical applications of magnetic carriers*. Plenum Press, New York.
Harada, Y., O. Ohara, A. Takatsuki, H. Itoh, N. Shimamoto, and K. Kinosita, Jr. 2001. Direct observation of DNA rotation during transcription by Escherichia coli RNA polymerase. *Nature* 409:113-115.
Harrison, D. J., K. Fluri, K. Seiler, Z. Fan, C. S. Effenhauser, and A. Manz. 1993. Micromachining a miniaturized capillary electrophoresis-based chemical analysis system on a chip. *Science* 261:895-897.
Johnson, K. S., M. Drndic, J. H. Thywissen, G. Zabow, R. M. Westervelt, and M. Prentiss. 1998. Atomic deflection using an adaptive microelectromagnet mirror. *Phys. Rev. Lett.* 81:1137-1141.
Kittel, C., and H. Kroemer. 1980. *Thermal physics*. W. H. Freeman, San Francisco.
Lee, C. S., H. Lee, and R. M. Westervelt. 2001. Microelectromagnets for the control of magnetic nanoparticles. *Appl. Phys. Lett.* 79:3308-3310.
Mehta, A. D., M. Rief, J. A. Spudich, D. A. Smith, and R. M. Simmons. 1999. Single-molecule biomechanics with optical methods. *Science* 283:1689-1695.
Morgan, H., M. P. Hughes, and N. G. Green. 1999. Separation of submicron bioparticles by dielectrophoresis. *Biophys. J.* 77:516-525.
Neher, E., and B. Sakmann. 1976. Single-channel currents recorded from membrane of denervated frog muscle fibres. *Nature* 260:799-802.
Neumann, D., O. Kohlbacher, H. P. Lenhof, and C. M. Lehr. 2002. Lectin-sugar interaction. Calculated versus experimental binding energies. *Eur. J. Biochem.* 269:1518-1524.

Perkins, T. T., D. E. Smith, and S. Chu. 1994. Direct observation of tube-like motion of a single polymer chain. *Science* 264:819-822.
Quake, S. R., and A. Scherer. 2000. From micro- to nanofabrication with soft materials. *Science* 290:1536-1540.
Radbruch, A., B. Mechtold, A. Thiel, S. Miltenyi, and E. Pfluger. 1994. High-gradient magnetic cell sorting. *Methods Cell Biol.* 42 Pt B:387-403.
Reif, F. 1965. *Fundamentals of statistical and thermal physics*. McGraw-Hill, New York.
Rembaum, A., and W. J. Dreyer. 1980. Immunomicrospheres: reagents for cell labeling and separation. *Science* 208:364-368.
Ryu, W. S., R. M. Berry, and H. C. Berg. 2000. Torque-generating units of the flagellar motor of Escherichia coli have a high duty ratio. *Nature* 403:444-447.
Sharon, N., and H. Lis. 1972. Lectins: cell-agglutinating and sugar-specific proteins. *Science* 177:949-959.
Svoboda, K., C. F. Schmidt, D. Branton, and S. M. Block. 1992. Conformation and elasticity of the isolated red blood cell membrane skeleton. *Biophys. J.* 63:784-793.
Whitesides, G. M., E. Ostuni, S. Takayama, X. Jiang, and D. E. Ingber. 2001. Soft lithography in biology and biochemistry. *Annu Rev Biomed Eng* 3:335-373.
Ziemann, F., J. Radler, and E. Sackmann. 1994. Local measurements of viscoelastic moduli of entangled actin networks using an oscillating magnetic bead micro-rheometer. *Biophys. J.* 66:2210-2216.

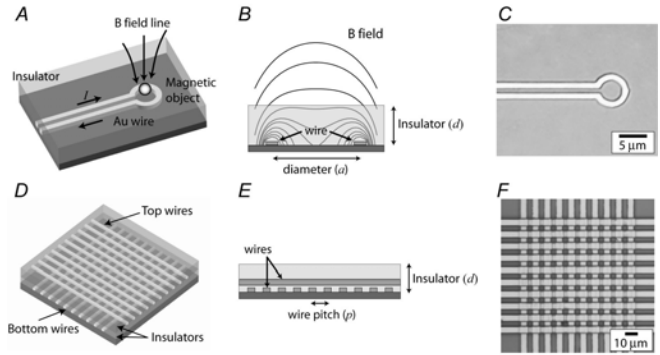

**Figure 1** (*A*) Schematic diagram of a microelectromagnet ring trap. The current *I* generates a magnetic field maximum on the surface of the insulator, where magnetic objects will be trapped. (*B*) Schematic cross section of a ring trap of diameter *a* with contours of the magnetic field magnitude. To create the maximum in the magnetic field magnitude at the center of the ring, the insulator thickness (*d*) is ≥ 0.32*a*. (*C*) Micrograph of a ring trap of radius 5 μm fabricated on a silicon/silicon oxide substrate. The thickness of an insulating layer is 2 μm. (*D*) Schematic diagram of a microelectromagnet matrix. The matrix consists of two arrays of straight wires aligned perpendicular to each other. The first set of conducting wires (bottom wires) is covered with an insulating layer, on which the second set of conducting wires (top wires) and an additional insulating layer are fabricated. The level of the current in each wire is individually controlled. (*E*) Schematic cross section of a matrix with the wire pitch *p*. The total thickness of the insulating layer (*d*) is $\geq 3p/(2^{3/2}\pi)$ to prevent the formation of the magnetic field maximum right on top of wires. (*F*) Micrograph of a fabricated matrix with 10 wires in each layer (a 10×10 wire matrix) with a wire spacing 8 μm. The distance between the substrate and the top surface of the device is 4 μm.

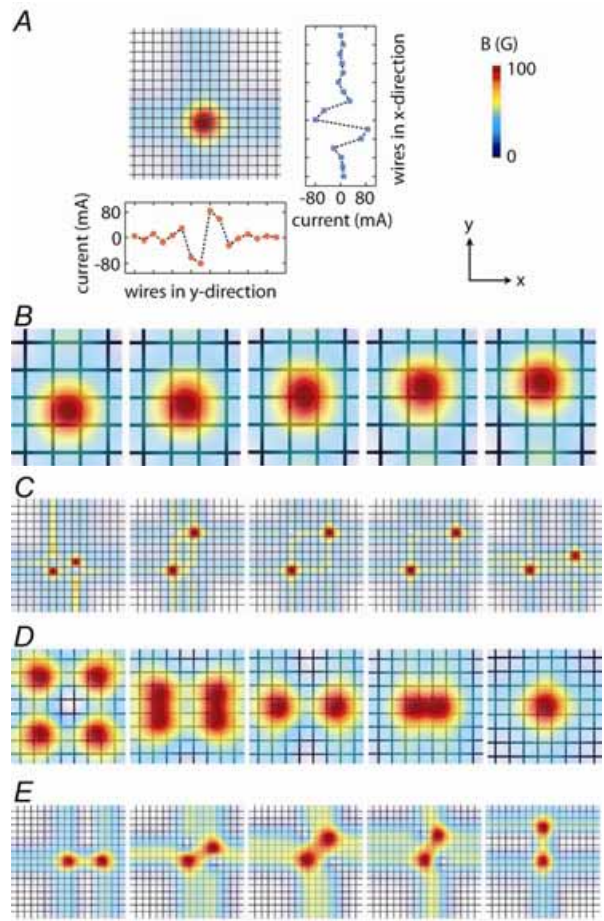

**Figure 2** Simulated magnetic field patterns produced by a 16×16 wire matrix. The wire pitch and the total thickness of insulator were 8 μm and 4 μm, respectively and the magnetic field magnitude was computed on the surface of the device. The solid lines show wire positions and the color bar scale corresponds to the magnetic field magnitude. (*A*) A single peak in the magnetic field magnitude with currents values in wires. Currents were optimized to generate a Gaussian shape peak (*B* = 10 mT). (*B*) By changing current distributions, the single peak was moved between two adjacent wires in steps less than the wire pitch. (*C*) Current control also allows simultaneous and independent movements of multiple peaks. In this example, two separate peaks were created and controlled: while one peak was fixed, the other was moved independently. (*D*) Four peaks moved independently and joined at one position. Reversing the sequence, this profile can be used for separations. (*E*) With one peak fixed at the center, the other peak circled around it, which can be used to rotate magnetically tagged biological samples.

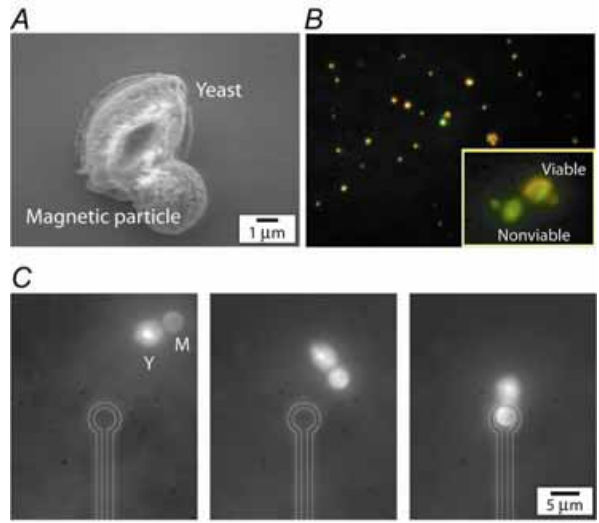

**Figure 3** (*A*) Scanning electron microscope image of a yeast cell (*Saccharnomyces cerevisiae*) bound to a magnetic particle. The particle was coated with Concanavalin-A which makes a specific binding to sugar molecules expressed on the wall of yeast cell. (*B*) Fluorescent microscope image of yeast samples stained with a two-color dye. Viable cells are exhibiting red intravacuolar structures whereas non viable cells are diffusively green (Inset). (*C*) Sequence of micrographs demonstrating a single cell trapping with a ring trap. Shortly after *I* = 50 mA was applied, generating a magnetic field peak *B* = 6 mT at the center of the ring, a single yeast cell with a magnetic particle was trapped and remained inside the trap. Y and M indicate yeast and magnetic particle, respectively.

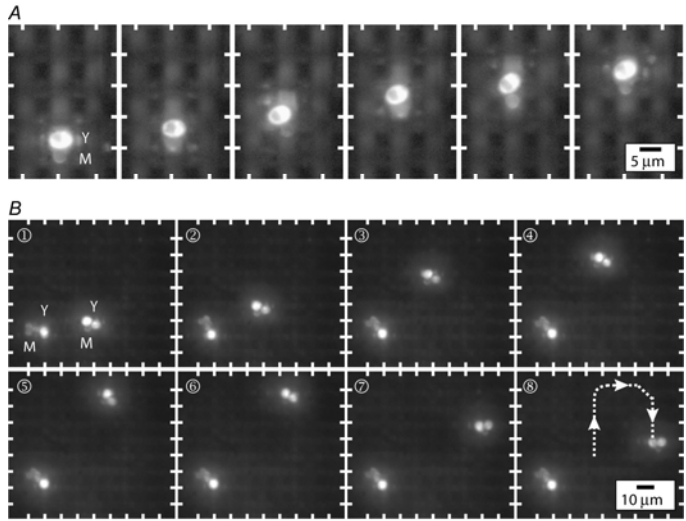

**Figure 4** Demonstration of trapping and transport of yeast samples by a 10×10 wire microelectromagnet matrix with the wire pitch 8 μm. Currents in all 20 wires were adjusted to generate desired magnetic field patterns. (*A*) A single magnetic field peak was created and moved continuously over two wires of the matrix, transporting a yeast cell by increments smaller than the wire spacing. (*B*) While a single cell with three magnetic particles was being held, a group of two cells with a magnetic particle was moved using the magnetic field pattern shown in Fig. 2 *B*. The matrix allows simultaneous and independent control of multiple samples without any modification to the structure of the device. White ticks show wire positions. Y and M indicate yeast and magnetic particle, respectively.

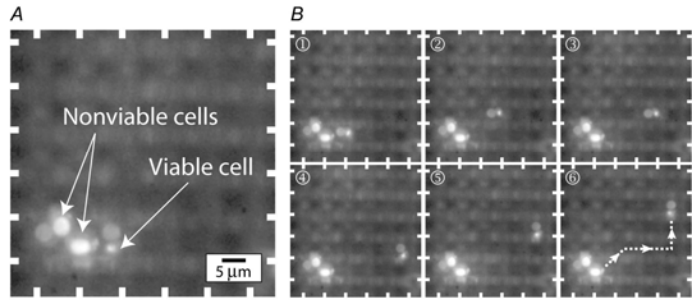

**Figure 5** Cell sorting operations with a microelectromagnet matrix. (*A*) Two nonviable cells and a viable cell were trapped with a single magnetic field peak. The viability of cells was identified by their fluorescent colors. (*B*) After initial trapping, the single peak was split into two smaller peaks. While the nonviable cells were fixed by a peak, the viable cell was separated by moving the other peak. Through the current control, optimized magnetic field patterns can be created to meet specific experimental needs. White ticks indicate wire positions.

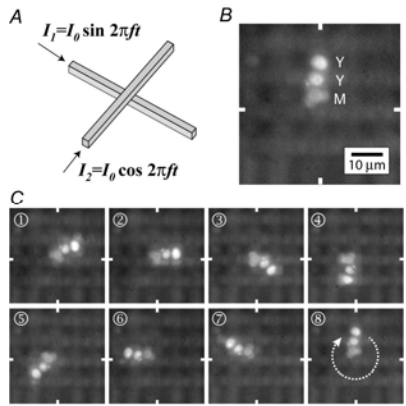

**Figure 6** Besides transporting target samples, the matrix can rotate or twist biological system using time-varying currents. (*A*) Schematic diagram for a rotating magnetic field generation. Two currents at the same frequency *f* but with phase difference 90° create a rotating magnetic field with the maximum in the field magnitude formed at the crossing point of two wires. (*B*) Two cells (Y) with three magnetic particles (M) were trapped on the crossing point of two wires. Ticks indicate the wires used for the field generation. (*C*) One complete revolution of the trapped cells. The maximum frequency for rotation was 2 Hz, which agrees well with the expected value 2.8 Hz.